\documentclass[twocolumn]{styles/aastex701}

\usepackage[utf8]{inputenc}
\usepackage{graphicx,float}
\usepackage[ruled,vlined]{algorithm2e}
\usepackage{algorithmic}
\usepackage{amsmath}
\usepackage{subcaption}
\usepackage{booktabs}
\usepackage{tabularx}
\usepackage{longtable}
\usepackage{threeparttable}

\usepackage{xcolor}
\usepackage{makecell}

\newcommand{\Gaia}{\textit{Gaia}}

\newcommand{\kmsec}{\mbox{km~s$^{\rm -1}$}}

\newcommand{\msun}{\mbox{$M_{\odot}$}}

\newcommand{\teff}{\mbox{$T_{\rm eff}$}}
\newcommand{\logg}{\mbox{log~{\it g}}}
\newcommand{\vt}{\mbox{$v_{\rm t}$}}

\newcommand{\feh}{\mbox{[Fe/H]}}
\newcommand{\ebv}{\mbox{$E(B-V)$}}

\newcommand{\Angstrom}{\text{\normalfont\AA}}

\renewcommand{\ion}[2]{#1\,\textsc{#2}}

\definecolor{BrickRed}{HTML}{F26035}

\usepackage{multirow}

\shorttitle{Sr and Ba yields of the First Generation(s) of stars}

\shortauthors{Hughes et al.}

\begin{document}

\title{Sr and Ba yields of the First Generation(s) of stars: Constraints from metal-poor stars
\footnote{This paper includes data gathered with the 6.5 meter Magellan Telescopes located at Las Campanas Observatory, Chile.}}

\author[0000-0002-7332-2751]{Sarah Hughes}
\affiliation{Department of Physics and Kavli Institute for Astrophysics and Space Research, Massachusetts Institute of Technology, Cambridge, MA 02139, USA}
\email{slhughes@mit.edu}
\correspondingauthor{Sarah Hughes}

\author[0000-0002-2139-7145]{Anna Frebel}
\affiliation{Department of Physics and Kavli Institute for Astrophysics and Space Research, Massachusetts Institute of Technology, Cambridge, MA 02139, USA}
\email{afrebel@mit.edu}

\author[0000-0002-4669-9967]{Xiaowei Ou}
\affiliation{Department of Astronomy, University of Virginia, Charlottesville, VA 22904, USA}
\affiliation{The NSF-Simons AI Institute for Cosmic Origins, USA}
\email{trg9de@virginia.edu}

\author[0000-0002-1462-0265]{Alexander Yelland}
\affiliation{Department of Physics and Kavli Institute for Astrophysics and Space Research, Massachusetts Institute of Technology, Cambridge, MA 02139, USA}
\email{ayelland@mit.edu}

\author[0009-0003-8577-9796]{Felicia Xiao} 
\affiliation{Department of Physics and Kavli Institute for Astrophysics and Space Research, Massachusetts Institute of Technology, Cambridge, MA 02139, USA}
\email{feli@mit.edu}

\author[0009-0007-0225-6519]{Jorian Benke} 
\affiliation{Department of Physics and Kavli Institute for Astrophysics and Space Research, Massachusetts Institute of Technology, Cambridge, MA 02139, USA}
\affiliation{Department of Aeronautics and Astronautics, Massachusetts Institute of Technology, Cambridge, MA 02139, USA}
\email{jbenke@mit.edu}

\author[0009-0005-7338-6594]{Kali Kraus} 
\affiliation{Department of Physics and Kavli Institute for Astrophysics and Space Research, Massachusetts Institute of Technology, Cambridge, MA 02139, USA}
\email{kmkraus@mit.edu}

\author[0000-0000-0000-0000]{Reidyn Wingate} 
\affiliation{Department of Physics and Kavli Institute for Astrophysics and Space Research, Massachusetts Institute of Technology, Cambridge, MA 02139, USA}
\email{raw123@mit.edu}

\author[0000-0001-9178-3992]{Mohammad K.\ Mardini}
\affiliation{Department of Physics and Kavli Institute for Astrophysics and Space Research, Massachusetts Institute of Technology, Cambridge, MA 02139, USA}
\email{mmardini@mit.edu}

\begin{abstract}

We present our chemical abundance analysis of ten new extremely metal-poor stars with $-4.05\leq\mbox{[Fe/H]}\leq-2.33$, based on high-resolution (R $\sim28,000$) Magellan/MIKE spectra. Eight of our stars have low heavy-element abundances of $\mbox{[Sr/H]}<-4.5$ and $\mbox{[Ba/H]}<-4.0$, making them Small Accreted Stellar System (SASS) stars. Four are hyper neutron-capture-element poor with $\mbox{[Sr/H]}<-5.0$, including Gaia DR3 5729400267359655680, which sets a new record for the lowest detected Sr abundance of $\mbox{[Sr/H]} =-6.4$. 
We identify four distinct [Sr/Ba] groups within the wider SASS star population which span a large range from $\mbox{[Sr/Ba]} =-2.0$ to +1.6, pointing to multiple types of progenitor events and different nucleosynthesis processes/sites.
{To explore the origins of this large [Sr/Ba] range, we adopt site-agnostic Sr yields of $\mbox{[Sr/H]}=-6$, $-5.75$, $-5.42$, and $-4.93$ for the four groups. Applying those yields suggests that the majority of SASS stars formed from gas enriched by $\sim$1-10 progenitor events, consistent with expectations from their extremely metal-poor nature. 
We thus attribute the [Sr/H] abundance scatter to intrinsic variations in the Sr yield per nucleosynthesis site/event. Our proposed Sr yields for each [Sr/Ba] group and associated nucleosynthesis origin are a reasonable and representative approximation, good to within a factor of a few, and can constrain future theoretical heavy element nucleosynthesis calculations in early core-collapse supernovae. }

\end{abstract}



\section{Introduction}

The most metal-poor stars display extremely low abundances of elements heavier than hydrogen and helium, making them ideal tracers of the first (Population III) stars and first galaxies, as well as the earliest nucleosynthesis events that set off the chemical evolution of the universe. 
{These rare stars are found in the galactic halo \citep{beers2005discovery, frebel2015near} and various dwarf satellite galaxies} \citep{frebel2010ApJ, VEN12, simon19} which allow a glimpse into the early formation process of the Milky Way and the associated star formation and element enrichment events. 

However, observing individual stars in ultra-faint dwarf galaxies {(UFDs; see \citealt{simon19} for a definition)} with high-resolution spectroscopy is notoriously difficult due to their large distances, sometimes out to a few hundreds of kiloparsecs into the halo. This renders only the brightest, cool red giants observable (with V $\sim17-19$), of which there often are only few available. Indeed, about a dozen UDFs have only three or fewer giants observed with high-resolution spectroscopy. Consequently, most UFDs do not have statistically well-characterized elemental abundance signatures. 

Given these limitations, studying neutron-capture elements is an alternative path to probing the onset of galaxy formation with stars that presumably formed in the first or earliest galaxies. A striking signature of stars found in UFDs are their low levels of neutron-capture element abundances (i.e., Sr and Ba), alongside their otherwise halo-like light element abundance ratios. Additionally, when considering the [Sr/Ba] ratio vs [Ba/Fe], UFD stars tend to form their own track, separate from that of metal-poor halo stars. This may indicate a limited enrichment history by the earliest SNe in each of these systems. Indeed, studies \citep{ji2019chemical} comparing $[\mathrm{Sr}/\mathrm{Ba}]$ ratios in UFDs have found they are consistently lower, by $\sim1$\,dex, compared to halo stars of the same metallicity \citep{frebel2010linking, frebel2010ApJ, frebel2014segue, frebel2018nuclei, ji2020detailed, sitnova2025}. 

However, a small group of halo stars does share the same UFD characteristics of low neutron-capture element abundances, and they overlay with the UFD track. {In this paper, we thus study these halo stars given that they curiously resemble the current UFD stars. Given their low metallicity and likely accretion origin, their chemical abundance patterns are poised to trace the earliest galaxy formation processes and could provide information on how many supernovae (SNe) may have contributed heavy elements to the original host systems.} 

These stars also probe the poorly understood sources of heavy neutron-capture elements in the early universe. A main source is the rapid neutron-capture process (r-process), and neutron star (NS) mergers are currently the only confirmed astrophysical site  \citep{pian2017spectroscopic, abbott2017gw170817}. However, NS--NS mergers alone cannot reproduce the r-process element abundance trends observed across the Milky Way \citet{cote2019}, as it would require such events to occur in nearly every small early galaxy on rapid timescales. An additional prompt r-process site, active at low metallicity, must account for roughly 50\% of early r-process element production. Alternative sites include some rare types of core-collapse supernovae, such magnetorotational supernovae \citep{nishimura15, Mosta18, yong2021nature} and collapsars \citep{siegel19,issa25}. Other sources of neutron-capture elements include a limited r-process \citep{wanajo2011}, and the weak s-process operating in massive stars \citep{Cescutti2013, choplin2017}.

{To systematically build a sample of these low Sr halo stars,} we have previously explored a sample of six extremely metal-poor stars with low neutron-capture element abundances (e.g., [Sr/H] $<-4.5$), three of which proved reminiscent of UFD stars and also having accretion-origin kinematics \citep{Andales}. These stars were labeled as ``Small Accreted Stellar System'', or SASS, stars. By extending the search to the \mbox{JINAbase} compilation \citep{abohalima2018jinabase}, \citet{Andales} identified an additional 61 SASS stars with the same heavy element and accretion characteristics.

The present study expands this exploration by contributing detailed chemical abundances for 10 new extremely metal-poor stars with extremely low neutron-capture element abundances, of which eight are SASS stars and four are hyper neutron-capture-element poor with [Sr/H] $<-5.0$. We use the entirety of the SASS sample to estimate the numbers of first/early supernovae that may have enriched their original hosts systems, what heavy element nucleosynthesis yields they may have had, and if there is any intrinsic scatter. Altogether, this allows us to better characterize the nature and birth environments of the oldest surviving stars. 

\section{Observations and Measurements}

\subsection{Target selections and existing observations}

The present study continues our effort to identify the most chemically primitive stars in the Milky Way, as determined from metallicity-sensitive photometry derived from Gaia XP spectrophotometry  (e.g., \citealt{Rix2022,Andrae2023,yao2024,Zhang2023}) along with the grids of synthetic photometry presented in \citet{Chiti21} and Mardini et al. (2026, in preparation). This catalog has yielded multiple discoveries, including the most metal-poor stars in the Large Magellanic Cloud down to [Fe/H] $-4.2$ \citep{Chiti24}, several ultra metal-poor stars \citep{mardini_srstar, Limberg25}, and the oldest and most metal-poor star known to host a transiting giant planet \citep{Simon26}.  

\begin{figure*}
\centering
    \includegraphics[width=\linewidth]{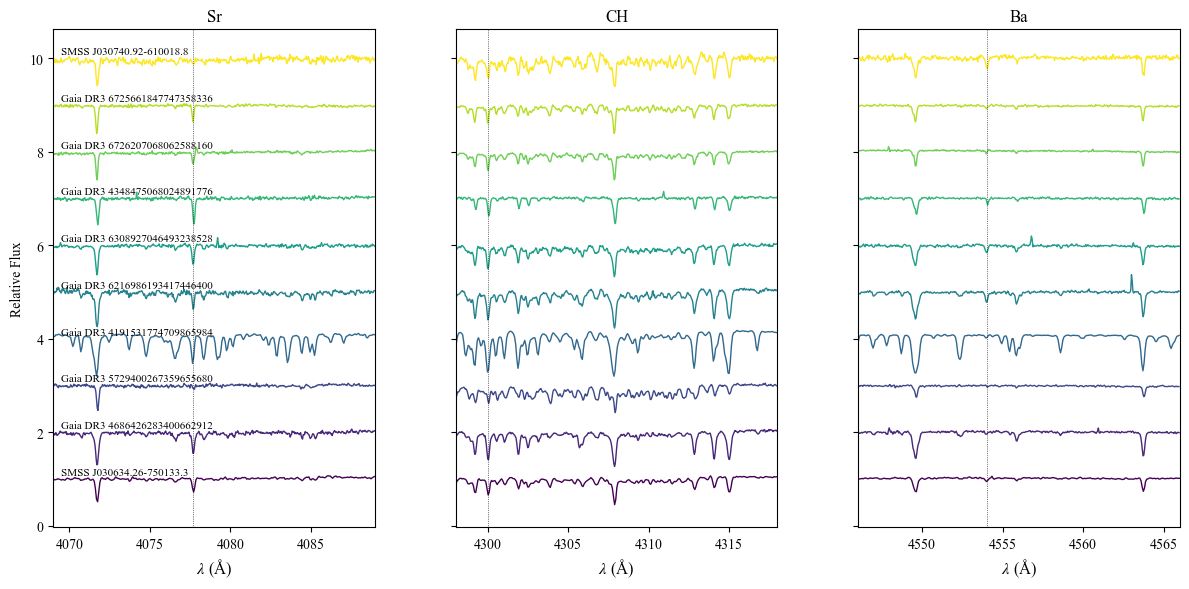} 
\caption{Spectra of our program stars around the Sr line at 4077\,{\AA}, the carbon G-band around 4300\,{\AA}, and the Ba line at 4554\,{\AA}. }
        \label{fig:spectralregions}
\end{figure*}

Eight of our program stars were initially selected from this catalog for high-resolution follow-up spectroscopy based on their low $\mbox{[Fe/H]}$ abundance estimates. In our sample of several hundred Gaia XP extremely metal-poor candidates for which we have already collected high-resolution spectra, we visually selected cool giant stars solely based on displaying a very weak or no Ba line at 4554\,{\AA}. This primarily visual selection resulted in a sample of eight extremely metal-poor stars for this study of halo stars with a high likelihood for extremely low Ba, and also Sr, abundances. As shown in \citet{Andales}, low Sr and Ba abundances preferentially select for likely long-accreted dwarf galaxy stars. We later supplemented the sample with two additional stars from \cite{Yong2021} (SMSS J030740.92$-$610018.8 and SMSS J030634.26$-$750133.3) with additional high-resolution spectra from our search for the most metal-poor stars.

\begin{deluxetable*}{lrrrrrrrrrr}
\tablecaption{Observing details and heliocentric velocities \label{tab:observations}}
\tablecolumns{11}
\tablewidth{0pt}
\tabletypesize{\scriptsize}
\tablehead{
\colhead{Star} & \colhead{RA} & \colhead{Dec} & \colhead{UT Dates} & \colhead{Slit} & \colhead{$t_{\text{exp}}$} & \colhead{$G$} & \colhead{$S/N$ at} & \colhead{$S/N$ at} & \colhead{$v_{\text{helio}}$} & \colhead{$v_{\text{helio, Gaia}}$} \\[-8pt]
\colhead{} & \colhead{[J2000]} & \colhead{[J2000]} & \colhead{} & \colhead{} & \colhead{[s]} & \colhead{[mag]} & \colhead{4000\,\AA} & \colhead{4500\,\AA} & \colhead{[km\,s$^{-1}$]} & \colhead{[km\,s$^{-1}$]}
}
\startdata
{Gaia DR3 4686426283400662912}  & 01:23:44.37 & $-$73:32:02.53 & 2024 07 02 & 0\farcs7 &  900 & 12.2  & 105 & 171 &  $-$56.3 &  $-$57.8 \\
SMSS J030634.26$-$750133.3 & 03:06:34.27 & $-$75:01:33.20 & 2017 10 08 & 1\farcs0 & 5100 & 13.7  &  60 &  96 &   145.9  &   145.8 \\
SMSS J030740.92$-$610018.8 & 03:07:40.95 & $-$61:00:18.76 & 2017 10 08 & 1\farcs0 & 7200 & 14.75 &  27 &  55 &   297.5 &   297.8 \\
{Gaia DR3 5729400267359655680} & 09:09:21.71 & $-$16:38:21.16 & 2025 03 11 & 0\farcs7 & 5400 & 13.9  &  59 &  98 &   125.9  &   120.5  \\
{Gaia DR3 6216986193417446400}  & 14:36:26.54 & $-$31:27:26.22 & 2025 03 11 & 0\farcs7 & 1200 & 13.3  &  32 &  54 &   197.7  &   197.9  \\
{Gaia DR3 6308927046493238528}        & 15:05:02.13 & $-$13:44:20.12 & 2025 03 25 & 1\farcs0 & 1800 & 13.7  &  30 &  45 & $-$242.3 & $-$237.7 \\
{Gaia DR3 4348475068024891776}        & 16:10:32.29 & $-$08:15:38.40 & 2025 03 11 & 0\farcs7 & 1200 & 13.2  &  54 &  79 & $-$141.7 & $-$142.3 \\
{Gaia DR3 6725661847747358336}        & 18:09:43.62 & $-$40:52:07.76 & 2025 03 11 & 0\farcs7 & 1200 & 10.7  &  73 & 102 &   464.2  &   466.0  \\
{Gaia DR3 6726207068062588160}        & 18:21:50.82 & $-$41:03:33.57 & 2025 03 11 & 0\farcs7 & 1800 & 13.2  &  94 & 111 &    24.1  &    24.3  \\
{Gaia DR3 4191531774709865984}  & 20:05:50.15 & $-$09:27:47.20 & 2024 07 02 & 0\farcs7 &  600 & 12.0  &  56 &  65 &   104.5  &   105.1  \\
\enddata
\tablecomments{The last two columns list the heliocentric velocities measured in this work and those adopted from \citet{Gaia_DR3}, respectively.}
\end{deluxetable*}

\subsection{High-resolution spectroscopy}

All ten stars were observed at Las Campanas Observatory in 2017, 2024, and 2025 using the MIKE instrument \citep{Bernstein03} on the Magellan-Clay telescope using slit sizes of $0\farcs7$ and $1\farcs0$. The resulting spectra have a resolving power of $\sim$28,000 in the red wavelength and $\sim$35,000 in the blue, and $\sim$22,000, and $\sim$28,000 respectively. The complete wavelength coverage of our spectra is $3330$-$9400$\,{\AA}. Exposure times ranged from $\sim$250 to 5500 seconds.

We used the CarPy MIKE\footnote{\url{https://code.obs.carnegiescience.edu/pipelines/mike}} pipeline to reduce all spectra \citep{kelson2003optimal}. We then normalized and stitched all blue and red echelle orders together before Doppler correcting the spectra through cross-correlation against the standard metal-poor star HD122563 in the region of the Ca triplet lines around 8500\,{\AA}. {Figure~\ref{fig:spectralregions} shows three selected spectral regions of the final spectra of all our stars: around the Sr line at 4077\,{\AA}, the CH G-band around 4300\,{\AA}, and the Ba line at 4554\,{\AA}.
Observational details and radial velocity measurements are listed in Table~\ref{tab:observations}. }

{Heliocentric velocities were determined with standard python packages (\textsc{numpy}: \citealt{numpy}; \textsc{scipy}: \citealt{scipy}), with uncertainties estimated from a Gaussian fit to the cross-correlation function peak. Associated uncertainties range from 0.3 to 1.2\,$\mathrm{km}$, with about half of them being at 0.5\,$\mathrm{km}$, while MIKE spectra typically have systematic uncertainties of 1-2\,$\mathrm{km}$. An overall level of 1-2\,$\mathrm{km}$ uncertainty suggests good agreement with the Gaia \citep{Gaia_DR3} velocities (see Table~\ref{tab:observations}). Two stars have discrepancies of 4-5\,$\mathrm{km}$ which might indicate binarity.}

{Two of our stars, SMSS J030740.92$-$610018.8 {(Gaia DR3~4723249928541065216)} and SMSS J030634.26$-$750133.3 {(Gaia DR3~4639676943066939776)}, were analyzed by \citet{Yong2021}. They measured $300.9$ and $143.2$\,km\,s$^{-1}$, respectively, which is broadly consistent to with in our uncertainties. Finally, we note that one star, Gaia DR3 6725661847747358336, has a unusually large radial velocity of $464.2$\,km\,s$^{-1}$ but which is confirmed by \textit{Gaia}'s value of 466.0\,km\,s$^{-1}$.}

\section{Stellar Parameter and Abundance Determination}

Equivalent width (EW) were measured of various absorption lines in the range of $\sim3500$-$7000\,{\Angstrom}$ to derive chemical abundances of $\alpha$-elements (Mg, Si, Ca, Ti), and iron-peak elements (Fe, Cr I, Ni, Zn). Fe lines with reduced equivalent width of $>-4.5$ are exclude to ensure that all lines are within the linear regime of the curve of growth. 
For carbon and the neutron-capture elements Sr and Ba, we used a spectrum synthesis approach. We used the linelist from \citet{Roederer18_lines} and solar abundances from \citet{asplund09}. For all line measurements we manually ensure correct continuum placement and good fitting of the Gaussian, especially in lower $S/N$ spectra. 

\begingroup
\setlength{\tabcolsep}{2pt}
\begin{deluxetable*}{lrrrrrrrrrrrrrrr}
\tabletypesize{\tiny}
\caption{\label{tab:kinematics}Stellar parameters and calculated orbital parameters of our sample stars, with energies, and stellar actions. Units are K, cgs, dex, km\,s$^{-1}$, {km\,s$^{-1}$}, {mas}, {mas}, kpc, $10^5$\,km$^2$\,s$^{-2}$, and kpc\,km\,s$^{-1}$ respectively. }
\tablehead{
  \colhead{Star} &
  \colhead{$T_{\text{eff}}$} &
  \colhead{$\log$ g} &
  \colhead{[Fe/H]} &
  \colhead{$v_{\text{mic}}$} &
  \colhead{{RV}} &
    \colhead{{$\pi$}} &
  \colhead{{$\pi_{\rm Corr}$}}&
  \colhead{$r_{\rm apo}$} &
  \colhead{$r_{\rm peri}$} &
  \colhead{$Z_{\rm max}$} &
  \colhead{$e$} &
  \colhead{$E$} &
  \colhead{$L_z$} &
  \colhead{$J_r$} &
  \colhead{$J_z$}
}
\startdata
{Gaia DR3 4686426283400662912} & 4462 & 0.68 & $-3.16$ & 2.38 & {$-62.5 \pm 0.9$} &  {$0.099\pm0.009$} &{0.102} & {$15.0\pm2.1$} & {$9.5\pm0.6$} & {$10.9\pm1.9$} & {$0.22\pm0.04$} & {$-$0.75$\pm0.04$} & {$1776.8\pm100.6$} & {$87.6\pm40.3$} & {$786.4\pm112.4$} \\
SMSS J030634.26$-$750133.3 & 5148 & 2.22 & $-2.98$ & $1.44$ & {$150.9 \pm 0.3$} &{$0.188\pm0.013$} &{0.220} & {$8.9\pm0.2$} & {$2.5\pm0.6$} &  {$5.5\pm0.6$} &  {$0.57\pm0.07$} & {$-$1.04$\pm0.02$} & {$-$444.3$\pm75.2$} & {$242.6\pm50.6$} & {$305.7\pm62.1$} \\
SMSS J030740.92$-$610018.8 & 5014 & 1.96 & $-3.09$ & 1.88 & {$303.45 \pm 0.6$} &{$0.077\pm0.015$} &{0.110} & {$13.00\pm1.20$} & {$6.13\pm0.36$} & {$9.53\pm0.34$} & {$0.36\pm0.02$} & {$-$0.84$\pm0.03$} & {$-$907.82$\pm156.44$} &  {$162.01\pm30.42$} & {$646.31\pm60.45$} \\
{Gaia DR3 5729400267359655680} & 4648 & 1.22 & $-4.05$ & 1.99 & {$136.0 \pm 0.7$} &{$0.018\pm0.017$} &{0.058} & {$18.70\pm2.26$} & {$1.35\pm0.79$} & {$5.63\pm0.67$} & {$0.87\pm0.05$} & {$-$0.79$\pm0.05$} & {$45.25\pm390.2$} & {$1209.9\pm107.6$} & {$80.2\pm16.7$} \\
{Gaia DR3 6216986193417446400} & 4237 & 0.13\rlap{\tablenotemark{a}} & $-3.18$ & 2.66 & {$173.3 \pm 0.5$} &{$0.005\pm0.017$} &{0.048} & {\nodata} & {\nodata} &{\nodata} & {\nodata} & {\nodata} &{\nodata} & {\nodata} & {\nodata} \\
{Gaia DR3 6308927046493238528} & 4810 & 1.29 & $-3.17$ & 1.86 & {$-263.3 \pm 0.3$} &{$0.048\pm0.022$} &{0.091}\rlap{\tablenotemark{b}} & {\nodata}  & {\nodata} &{\nodata}  & {\nodata}  & {\nodata}  & {\nodata}  & {\nodata}  & {\nodata} \\
{Gaia DR3 4348475068024891776} & 4990 & 1.64 & $-3.50$ & 1.8 & {$-169.7 \pm 1.0$} &{$0.158\pm0.016$} &{0.200} & {$45.24\pm11.87$} & {$4.24\pm0.31$} & {$36.17\pm10.27$} & {$0.82\pm0.05$} & {$-$0.48$\pm0.07$} & {$1059.7\pm78.1$} & {$2349.8\pm816.4$} &  {$837.1\pm39.6$} \\
{Gaia DR3 6725661847747358336} & 4993 & 1.89 & $-3.45$ & 1.77 & {$436.0 \pm 1.1$} &{$0.170\pm0.017$} &{0.212} & {$42.71\pm3.1$} & {$2.3\pm0.3$} & {$8.6\pm1.5$} &  {$0.90\pm0.01$} & {$-$0.5$\pm0.02$} & {$-$975.6$\pm49.3$} & {$2683.9\pm189.9$} &  {$61.2\pm9.2$} \\
{Gaia DR3 6726207068062588160} & 5108 & 1.90 & $-3.47$ & 1.60 & {$-3.75 \pm 1.0$} &{$0.156\pm0.014$} &{0.199} & {$3.8\pm0.2$} & {$1.6\pm0.3$} & {$1.2\pm0.1$} &  {$0.41\pm0.09$} & {$-$1.44$\pm0.01$} & {$-$339.9$\pm39.3$} & {$62.3\pm31.3$} & {$66.2\pm13.2$} \\
{Gaia DR3 4191531774709865984} & 4378 & 0.64 & $-2.33$ & 2.12 & {$94.5 \pm 0.2$} &{$0.113\pm0.013$} &{0.143} & {$7.6\pm0.4$} & {$3.7\pm0.4$} & {$3.7\pm0.4$} & {$0.34\pm0.06$} & {$-$1.1$\pm0.01$} &  {$846.2\pm76.8$} & {$94.9\pm33.9$} & {$207.5\pm19.0$} \\
\enddata
\tablenotetext{a}{Due to a large parallax uncertainty, this $\log$ g is based on MIST isochrones.}
\tablenotetext{b}{{The action integrals did not converge for this star.}}
\end{deluxetable*}
\endgroup

{Stellar parameters were determined largely following \cite{ji2020s5}. We summarize the main steps here and present our results in Table~\ref{tab:kinematics}. 
Photometric effective temperatures, \teff\, were derived using \Gaia\ DR3 $G$-$RP$ colors \citep{Gaia_DR3} with color-[Fe/H]-\teff\ relations from \citet{Mucciarelli2021}.  
We calculated the unextincted magnitudes using the extinction law provided by \Gaia \footnote{\url{https://www.cosmos.esa.int/web/gaia/edr3-extinction-law}} and the extinction map (\ebv) from \cite{schlafly2011}.
[Fe/H] was initially set to be the Fe\,\textsc{i} abundance and then iterated until the final photometric stellar parameters stabilized for each star.}

{Specifically, we derive \logg\ using fundamental relations including the definition of surface gravity and the Stefan-Boltzmann equation:}

{\begin{equation}
\begin{aligned}
    \log g~=~4\log(\teff/T_{\rm{eff,\odot}})+\log(M/\msun)+\log g_{\odot} \\
    +0.4(BC_G+m_G-5\log(d)+5-M_{\text{bol},\odot}),
\end{aligned}
\end{equation}}

{For the fundamental relations, we assume the mass of the star to be $0.75 \pm 0.1$\,\msun\ for typical red giant stars, adopt the bolometric correction in the $G$~band from \citet{casagrande18}. We take $M_{\text{bol},\odot}$, $T_{\text{eff},\odot}$, and $\log g_{\odot}$ as the solar bolometric magnitude, effective temperature, and surface gravity, taken to be $4.75$, $5780$\,K, and $4.44$, respectively. Regarding distances, resolved parallax measurement are available for all but one star in our sample (see below). We calculate the distance in pc following \citet{Mardini2022} from the inverse parallax obtained from \Gaia\ (after correcting for zero point bias following \citealt{Lindegren_Parallax_2021}).}

{Among these input properties, $BC_G$ implicitly depends on the stellar \teff, \logg, [Fe/H], and \ebv, while the derivation of $m_G$ depends on \ebv.
These dependencies again require the stellar parameter derivation to be iterated until the final photometric stellar parameters stabilized.
Furthermore, the uncertainties in \teff, $M$, $BC_G$, $m_G$, and $d$ are propagated through a Monte Carlo method, by resampling $10^4$ times every input, including uncertainties in \ebv\ and apparent $G$ magnitude.}

{Star Gaia DR3 621698619341744640 has an unresolved parallax of $\sim0.005\pm0.017$\,mas (before zero point bias correction) which prevents a reliable \logg\ determination from the fundamental relations. Instead, we use low-metallicity MIST isochrones of the red giant branch \citep{dotter2016mist}.
We assume a stellar age of $10.5\pm0.5$\,Gyr, but note that metallicity influences the position of the RGB much more significantly than age. 
We then interpolate the isochrones in accordance with a star's metallicity to obtain their \logg. The uncertainties in \feh\ and age are propagated through by Monte Carlo sampling \feh\ and age, regenerating the isochrones, and repeating the interpolation.}

{For Gaia DR3 621698619341744640, we thus adopt the MIST \logg\  of 0.13\,dex.
But we also obtain \logg\ for our entire sample for comparison. Results are consistent to within \mbox{$\sim0.14$}\,dex (driven in part by Gaia DR3 5729400267359655680 which also has a small parallax). As such, we do not separately report these values.}

After obtaining final values for the surface gravity and the effective temperature, we then iteratively determined \vt\ and \feh\ from Fe\,\textsc{i}, following  guesses from an initial spectroscopic stellar parameter analysis. 

Statistical uncertainties on \teff\ were determined by propagating input uncertainties (i.e., the \Gaia\ photometry, extinction correction, and \feh\ uncertainties) through the color-[Fe/H]-\teff\ relations.
These uncertainties are propagated to the \logg\ estimates, where both methods depends on \teff.
Additionally, depending on the estimate methods, uncertainties from (1) the stellar mass and stellar distances are factored in for the fundamental relations estimates and (2) the stellar age is factored in for the isochrone fitting estimates.
The \vt~statistical uncertainty was determined from the 1-$\sigma$ confidence interval in the fitted slope with respect to reduced equivalent widths. 
{
Specifically, the uncertainties are propagated through a Monte Carlo method, by resampling $10^4$ times every input.
}
Typical statistical uncertainties are $<10$\,K for \teff\ (dominated by uncertainties in [Fe/H]), $\sim0.15$\,dex for \logg,  
$\sim0.15$\,\kmsec\ for \vt, and 0.15\,dex for [Fe/H].
Systematic uncertainties on \teff\ and \logg\ arise from choices in the color-[Fe/H]-\teff\ relations and isochrones (bolometric corrections). 
For \teff, we adopt  70\,K  as given in \citet{Mucciarelli2021}, and $0.2$\,dex and $0.2$\,\kmsec\ for \logg\ and \vt, respectively, following assessments, e.g., in \citet{frebel13}.

To arrive at our total uncertainties, we add the systematic uncertainties in quadrature to the statistical uncertainties. For all stars, we find the typical total uncertainties in \teff, \logg, and \vt\ to be 
$\sim 71\,{\rm K}$, 
$\sim 0.26\,{\rm dex}$, and 
$\sim 0.25\,{\rm km\,s^{-1}}$, respectively.

When calculating the abundances, we used ATLAS stellar atmosphere models with $\alpha$-enhancement from \citet{castelli_kurucz} along with the most recent version of the MOOG spectrum synthesis code\footnote{\url{https://github.com/alexji/moog17scat}}, which includes Rayleigh scattering and assumes 1D parallel-plane geometry \citep{Sneden1973, Sobeck2011}, and local thermodynamic equilibrium (LTE). This is all incorporated in the custom software SMHR\footnote{\url{https://github.com/andycasey/smhr}} \citep{Casey2014, 2025smhr}, which wraps MOOG to build an interactive platform for the analysis. Table~\ref{tab:abundances} lists the final abundance measurements for our sample. 

{Regarding abundance uncertainties, they were calculated based on the standard deviations $\sigma$ of the individual lines measured for each element, and adopting small number statistics to compute the associated standard errors \citep{keeping62}. They are given in Table~\ref{tab:abundances}. Specifically, for elements with only a few lines available ($2 < N < 30$) the small N-corrections applied to the standard deviation account for the limited sampling and the increased statistical uncertainty. 
In cases where only one line measurement was available, such as Zn, we assume a minimum nominal error of $0.1$\,dex or 0.15\,dex, depending on the quality of our data.}

\section{Discussion of chemical abundance results and SASS classification}
\label{sec:abundance_results}

We present the results of our chemical abundance analysis for the ten stars, summarized in Table~\ref{tab:abundances}. Figures~\ref{fig:elements} and \ref{fig:SrH_BaH} show our sample in comparison with a set of metal-poor halo taken from JINAbase \citep{abohalima2018jinabase}\footnote{\url{https://jinabase.pythonanywhere.com}} and UFD stars and other literature stars (full details given table~2 in \citealt{yelland26}). In the following, we discuss the results of our abundance analysis for several element groups, and the classification of our objects as SASS stars. We note that we limited the elements commonly analyzed in other studies, including \cite{Andales}, to a subset of elements most relevant to our overall analysis and with a focus on Sr and Ba abundances.

\begin{deluxetable*}{lrrrrr@{\hspace{6pt}}l@{}rrrrr@{\hspace{6pt}}l@{}rrrrr}
\renewcommand{\arraystretch}{0.85}
\tablecaption{Stellar abundances for our sample of 10 stars \label{tab:abundances}}
\tablecolumns{18}
\tablewidth{0pt}
\tabletypesize{\tiny}
\setlength{\tabcolsep}{2pt}
\tablehead{
\colhead{Species} & \colhead{$N$} & \colhead{$\log\epsilon(X)$} & \colhead{$\sigma$} & \colhead{[X/H]} & \colhead{[X/Fe]} &
\colhead{} & \colhead{$N$} & \colhead{$\log\epsilon(X)$} & \colhead{$\sigma$} & \colhead{[X/H]} & \colhead{[X/Fe]} &
\colhead{} & \colhead{$N$} & \colhead{$\log\epsilon(X)$} & \colhead{$\sigma$} & \colhead{[X/H]} & \colhead{[X/Fe]}\\
\multicolumn{1}{c}{} & \multicolumn{5}{c}{{Gaia DR3 4686426283400662912}} &
\multicolumn{1}{c}{} &\multicolumn{5}{c}{{Gaia DR3 5729400267359655680}} &
\multicolumn{1}{c}{} &\multicolumn{5}{c}{{Gaia DR3 6216986193417446400}} \vspace{-0.12cm}
}
\startdata
\ion{CH}{} & 2 & 4.51 & 0.06 & $-3.92$ & $-0.75$ &
 & 2 & 4.88 & 0.14 & $-3.55$ & $0.51$ &
 & 2 & 4.54 & 0.15 & $-3.89$ & $-0.73$ \\
\ion{CH$_{\mathrm{corr}}$}{} & 2 & 5.25 & 0.06 & $-3.18$ & $-0.01$ &
 & 2 & 5.52 & 0.14 & $-2.91$ & $1.15$ &
 & 2 & 5.28 & 0.15 & $-3.15$ & $0.01$ \\
\ion{Mg}{i} & 9 & 5.26 & 0.04 & $-2.34$ & $0.83$ &
 & 2 & 4.30 & 0.10 & $-3.30$ & $0.76$ &
 & 5 & 4.82 & 0.15 & $-2.78$ & $0.39$ \\
\ion{Si}{i} & 2 & 4.98 & 0.05 & $-2.53$ & $0.64$ &
 & 2 & 4.46 & 0.35 & $-3.05$ & $1.01$ &
 & 2 & 4.85 & 0.29 & $-2.66$ & $0.50$ \\
\ion{Ca}{i} & 22 & 3.49 & 0.11 & $-2.85$ & $0.32$ &
 & 11 & 2.94 & 0.14 & $-3.4$ & $0.66$ &
 & 15 & 3.35 & 0.14 & $-2.99$ & $0.17$ \\
\ion{Ti}{i} & 8 & 2.04 & 0.03 & $-2.91$ & $0.26$ &
 & 6 & 1.11 & 0.12 & $-3.84$ & $0.21$ &
 & 19 & 1.62 & 0.11 & $-3.33$ & $-0.16$ \\
\ion{Ti}{ii} & 26 & 2.11 & 0.10 & $-2.84$ & $0.32$ &
 & 14 & 1.09 & 0.08 & $-3.86$ & $0.19$ &
 & 28 & 1.79 & 0.09 & $-3.16$ & $0.00$ \\
\ion{Cr}{i} & 7 & 2.20 & 0.07 & $-3.44$ & $-0.27$ &
 & 3 & 1.03 & 0.30 & $-4.61$ & $-0.55$ &
 & 8 & 2.13 & 0.12 & $-3.51$ & $-0.35$ \\
\ion{Fe}{i} & 146 & 4.33 & 0.18 & $-3.17$ & $0.00$ &
 & 100 & 3.44 & 0.02 & $-4.06$ & $0.00$ &
 & 142 & 4.34 & 0.17 & $-3.16$ & $0.00$ \\
\ion{Fe}{ii} & 14 & 4.33 & 0.13 & $-3.17$ & $-0.01$ &
 & 10 & 3.54 & 0.14 & $-3.96$ & $0.10$ &
 & 15 & 4.40 & 0.10 & $-3.10$ & $0.06$ \\
\ion{Ni}{i} & 7 & 2.79 & 0.10 & $-3.43$ & $-0.26$ &
 & 2 & 1.98 & 0.03 & $-4.24$ & $-0.18$ &
 & 11 & 3.10 & 0.11 & $-3.12$ & $0.05$ \\
\ion{Zn}{i} & 3 & 1.58 & 0.19 & $-2.98$ & $0.19$ &
 & 1 & $<1.45$ & \nodata & $<-3.11$ & $<0.95$ &
 & 1 & 1.28 & 0.10 & $-3.28$ & $-0.12$ \\
\ion{Sr}{ii} & 2 & $-2.26$ & 0.07 & $-5.13$ & $-1.96$ &
 & 2 & $-3.48$ & 0.10 & $-6.35$ & $-2.29$ &
 & 2 & $-3.02$ & 0.02 & $-5.89$ & $-2.73$ \\
\ion{Ba}{ii} & 2 & $-3.94$ & 0.49 & $-6.12$ & $-2.97$ &
 & 1 & $-3.78$ & 0.10 & $-5.96$ & $-1.91$ &
 & 3 & $-3.16$ & 0.05 & $-5.34$ & $-2.14$ \\
\tableline\\
\multicolumn{1}{c}{} & \multicolumn{5}{c}{Gaia DR3 6308927046493238528} &
\multicolumn{1}{c}{} & \multicolumn{5}{c}{Gaia DR3 4348475068024891776} &
\multicolumn{1}{c}{} & \multicolumn{5}{c}{Gaia DR3 6725661847747358336} \\
\tableline
\ion{CH}{} & 2 & 5.17 & 0.06 & $-3.26$ & $-0.08$ &
 & 2 & 4.32 & 0.09 & $-4.11$ & $-0.55$ &
 & 2 & 5.05 & 0.02 & $-3.38$ & $0.06$ \\
\ion{CH$_{\mathrm{corr}}$}{} & 2 & 5.65 & 0.06 & $-2.78$ & $0.40$ &
 & 2 & 4.52 & 0.09 & $-3.91$ & $-0.35$ &
 & 2 & 5.07 & 0.02 & $-3.36$ & $0.08$ \\
\ion{Mg}{i} & 4 & 4.79 & 0.13 & $-2.80$ & $0.37$ &
 & 4 & 4.20 & 0.11 & $-3.40$ & $0.16$ &
 & 5 & 4.56 & 0.10 & $-3.04$ & $0.40$ \\
\ion{Si}{i} & 2 & 4.87 & 0.14 & $-2.64$ & $0.54$ &
 & 1 & 4.60 & 0.10 & $-2.91$ & $0.65$ &
 & 1 & 4.49 & 0.10 & $-3.02$ & $0.42$ \\
\ion{Ca}{i} & 13 & 3.39 & 0.07 & $-2.95$ & $0.23$ &
 & 11 & 2.86 & 0.09 & $-3.48$ & $0.08$ &
 & 10 & 3.14 & 0.08 & $-3.20$ & $0.24$ \\
\ion{Ti}{i} & 10 & 1.86 & 0.08 & $-3.09$ & $0.09$ &
 & 7 & 1.34 & 0.10 & $-3.61$ & $-0.05$ &
 & 7 & 1.73 & 0.14 & $-3.22$ & $0.22$ \\
\ion{Ti}{ii} & 14 & 1.98 & 0.09 & $-2.97$ & $0.20$ &
 & 16 & 1.42 & 0.11 & $-3.53$ & $0.03$ &
 & 15 & 1.69 & 0.11 & $-3.26$ & $0.19$ \\
\ion{Cr}{i} & 8 & 2.39 & 0.10 & $-3.25$ & $-0.07$ &
 & 3 & 1.19 & 0.17 & $-4.45$ & $-0.89$ &
 & 6 & 2.00 & 0.27 & $-3.64$ & $-0.20$ \\
\ion{Fe}{i} & 131 & 4.32 & 0.15 & $-3.18$ & $0.00$ &
 & 93 & 3.94 & 0.13 & $-3.56$ & $0.00$ &
 & 84 & 4.06 & 0.15 & $-3.44$ & $0.00$ \\
\ion{Fe}{ii} & 15 & 4.34 & 0.07 & $-3.16$ & $0.02$ &
 & 12 & 4.03 & 0.08 & $-3.47$ & $0.09$ &
 & 10 & 4.12 & 0.11 & $-3.38$ & $0.06$ \\
\ion{Ni}{i} & 4 & 3.24 & 0.08 & $-2.98$ & $0.20$ &
 & 2 & 2.27 & 0.14 & $-3.95$ & $-0.39$ &
 & 2 & 2.59 & 0.07 & $-3.63$ & $-0.19$ \\
\ion{Zn}{i} & 1 & 1.82 & 0.10 & $-2.74$ & $0.44$ &
 & 1 & $<1.44$ & \nodata & $<-3.12$ & $<0.43$ &
 & 1 & $<1.42$ & \nodata & $<-3.14$ & $<0.31$ \\
\ion{Sr}{ii} & 2 & $-1.85$ & 0.07 & $-4.72$ & $-1.54$ &
 & 2 & $-1.05$ & 0.10 & $-3.92$ & $-0.36$ &
 & 2 & $-2.03$ & 0.10 & $-4.90$ & $-1.46$ \\
\ion{Ba}{ii} & 2 & $-2.53$ & 0.11 & $-4.71$ & $-1.54$ &
 & 2 & $-2.60$ & 0.02 & $-4.78$ & $-1.22$ &
 & 2 & $-2.82$ & 0.17 & $-5.00$ & $-1.56$ \\
\tableline\\
\multicolumn{1}{c}{} & \multicolumn{5}{c}{Gaia DR3 6726207068062588160} &
\multicolumn{1}{c}{} & \multicolumn{5}{c}{SMSS J030634.26$-$750133.3} &
\multicolumn{1}{c}{} & \multicolumn{5}{c}{{Gaia DR3 4191531774709865984}} \\
\tableline
\ion{CH}{} & 2 & 5.32 & 0.06 & $-3.11$ & $0.36$ &
 & 2 & 5.31 & 0.04 & $-3.12$ & $-0.10$ &
 & 2 & 4.80 & 0.05 & $-3.63$ & $-1.30$ \\
\ion{CH$_{\mathrm{corr}}$}{} & 2 & 5.34 & 0.06 & $-3.09$ & $0.38$ &
 & 2 & 5.32 & 0.04 & $-3.11$ & $-0.09$ &
 & 2 & 5.59 & 0.05 & $-2.84$ & $-0.51$ \\
\ion{Mg}{i} & 5 & 4.78 & 0.09 & $-2.82$ & $0.65$ &
 & 5 & 4.78 & 0.08 & $-2.82$ & $0.20$ &
 & 2 & 4.90 & 0.06 & $-2.70$ & $-0.36$ \\
\ion{Si}{i} & 1 & 4.66 & 0.10 & $-2.85$ & $0.62$ &
 & 2 & 4.73 & 0.03 & $-2.78$ & $0.24$ &
 & 2 & 4.92 & 0.17 & $-2.59$ & $-0.26$ \\
\ion{Ca}{i} & 11 & 3.34 & 0.12 & $-3.00$ & $0.47$ &
 & 12 & 3.41 & 0.12 & $-2.93$ & $0.10$ &
 & 19 & 3.90 & 0.12 & $-2.44$ & $-0.10$ \\
\ion{Ti}{i} & 11 & 1.93 & 0.11 & $-3.02$ & $0.45$ &
 & 10 & 2.15 & 0.09 & $-2.80$ & $0.22$ &
 & 16 & 2.95 & 0.14 & $-2.00$ & $0.34$ \\
\ion{Ti}{ii} & 17 & 1.71 & 0.06 & $-3.24$ & $0.23$ &
 & 18 & 2.04 & 0.07 & $-2.91$ & $0.11$ &
 & 22 & 3.48 & 0.14 & $-1.47$ & $0.86$ \\
\ion{Cr}{i} & 3 & 1.78 & 0.25 & $-3.86$ & $-0.39$ &
 & 7 & 2.37 & 0.12 & $-3.27$ & $-0.25$ &
 & 12 & 3.59 & 0.12 & $-2.05$ & $0.28$ \\
\ion{Fe}{i} & 100 & 4.03 & 0.13 & $-3.47$ & $0.00$ &
 & 118 & 4.48 & 0.14 & $-3.02$ & $0.00$ &
 & 161 & 5.16 & 0.12 & $-2.34$ & $0.00$ \\
\ion{Fe}{ii} & 11 & 4.01 & 0.06 & $-3.49$ & $0.03$ &
 & 8 & 4.53 & 0.10 & $-2.97$ & $0.05$ &
 & 15 & 5.43 & 0.12 & $-2.07$ & $0.27$ \\
\ion{Ni}{i} & 2 & 2.78 & 0.09 & $-3.44$ & $0.03$ &
 & 4 & 3.24 & 0.16 & $-2.98$ & $0.04$ &
 & 17 & 4.35 & 0.11 & $-1.87$ & $0.47$ \\
\ion{Zn}{i} & 1 & 1.60 & 0.10 & $-2.96$ & $0.51$ &
 & 1 & 2.03 & 0.10 & $-2.53$ & $0.50$ &
 & 1 & 3.42 & 0.10 & $-1.14$ & $1.20$ \\
\ion{Sr}{ii} & 1 & $-2.16$ & 0.10 & $-5.03$ & $-1.56$ &
 & 2 & $-1.65$ & 0.10 & $-4.52$ & $-1.49$ &
 & 2 & $-1.18$ & 0.10 & $-4.05$ & $-1.72$ \\
\ion{Ba}{ii} & 2 & $-2.83$ & 0.01 & $-5.01$ & $-1.54$ &
 & 2 & $-2.48$ & 0.04 & $-4.66$ & $-1.63$ &
 & 3 & $-3.46$ & 0.18 & $-5.65$ & $-3.32$ \\
\tableline\\
\multicolumn{6}{c}{} &
\multicolumn{1}{c}{} & \multicolumn{5}{c}{SMSS J030740.92$-$610018.8} &
\multicolumn{6}{c}{} \\
\cline{7-12}
\multicolumn{6}{c}{} &
\ion{CH}{} & 2 & 5.45 & 0.04 & $-2.98$ & $0.11$ &
\multicolumn{6}{c}{} \\
\multicolumn{6}{c}{} &
\ion{CH$_{\mathrm{corr}}$}{} & 2 & 5.47 & 0.04 & $-2.96$ & $0.13$ &
\multicolumn{6}{c}{} \\
\multicolumn{6}{c}{} &
\ion{Mg}{i} & 5 & 5.25 & 0.08 & $-2.35$ & $0.73$ &
\multicolumn{6}{c}{} \\
\multicolumn{6}{c}{} &
\ion{Si}{i} & 2 & 5.64 & 0.03 & $-1.87$ & $1.22$ &
\multicolumn{6}{c}{} \\
\multicolumn{6}{c}{} &
\ion{Ca}{i} & 13 & 3.62 & 0.12 & $-2.72$ & $0.37$ &
\multicolumn{6}{c}{} \\
\multicolumn{6}{c}{} &
\ion{Ti}{i} & 7 & 2.05 & 0.09 & $-2.90$ & $0.19$ &
\multicolumn{6}{c}{} \\
\multicolumn{6}{c}{} &
\ion{Ti}{ii} & 18 & 2.03 & 0.07 & $-2.92$ & $0.17$ &
\multicolumn{6}{c}{} \\
\multicolumn{6}{c}{} &
\ion{Cr}{i} & 4 & 2.00 & 0.15 & $-3.64$ & $-0.55$ &
\multicolumn{6}{c}{} \\
\multicolumn{6}{c}{} &
\ion{Fe}{i} & 85 & 4.41 & 0.17 & $-3.09$ & $0.00$ &
\multicolumn{6}{c}{} \\
\multicolumn{6}{c}{} &
\ion{Fe}{ii} & 11 & 4.51 & 0.11 & $-2.99$ & $0.10$ &
\multicolumn{6}{c}{} \\
\multicolumn{6}{c}{} &
\ion{Ni}{i} & 3 & 3.16 & 0.16 & $-3.06$ & $0.03$ &
\multicolumn{6}{c}{} \\
\multicolumn{6}{c}{} &
\ion{Zn}{i} & 1 & $<1.64$ & \nodata & $<-2.92$ & $<0.17$ &
\multicolumn{6}{c}{} \\
\multicolumn{6}{c}{} &
\ion{Sr}{ii} & 2 & $-2.57$ & 0.10 & $-5.44$ & $-2.36$ &
\multicolumn{6}{c}{} \\
\multicolumn{6}{c}{} &
\ion{Ba}{ii} & 2 & $-1.90$ & 0.04 & $-4.08$ & $-1.00$ & \multicolumn{6}{c}{} \\
\cline{7-12}
\enddata
\tabletypesize{\small}
\end{deluxetable*}

\subsection{Light, iron-peak element abundances and abundance uncertainties}

\begin{figure*}
    \centering
    \includegraphics[width=\linewidth]{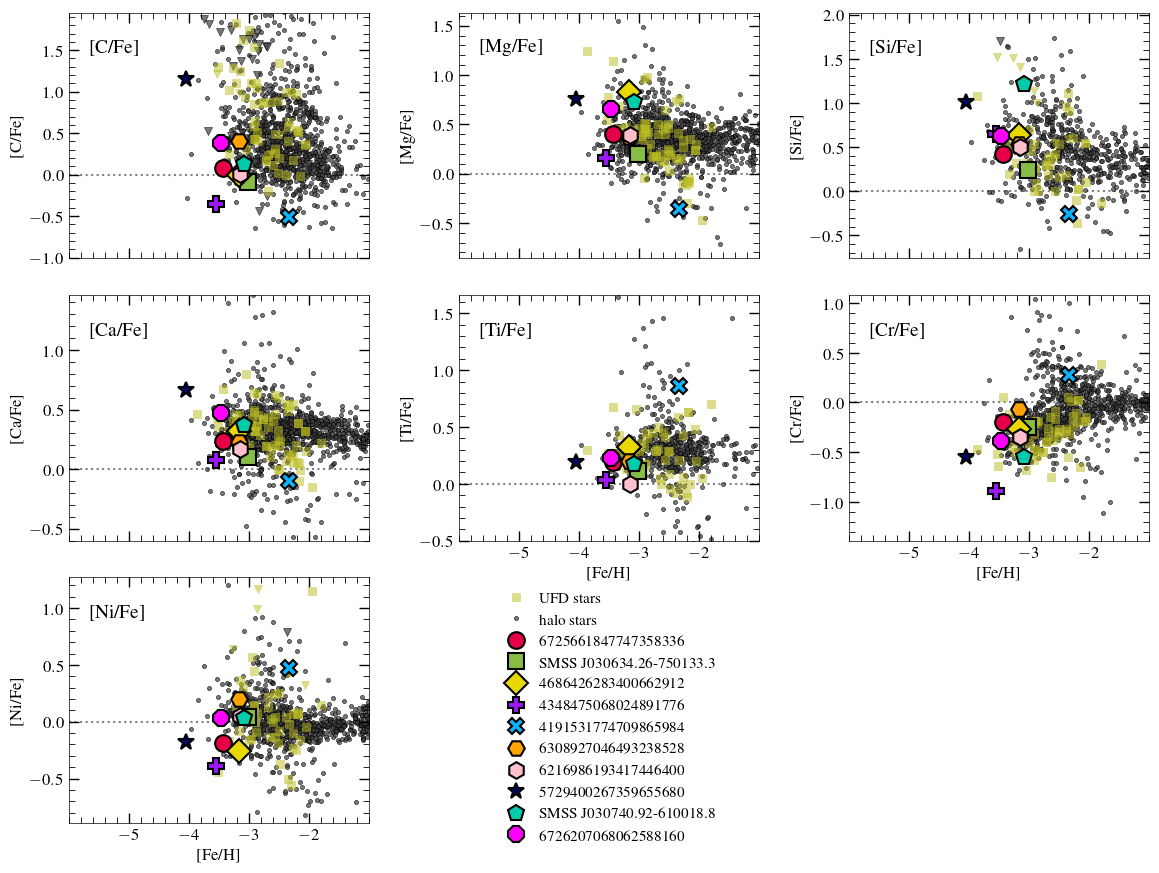}
    \caption{Abundances of light and iron-peak elements of our sample stars. For comparison, we also show metal-poor UFD and halo stars, represented by olive squares and gray points, respectively (taken from \citealt{Andales} and JINAbase \cite {Aoki2013a, Barklem2005b, Cohen2013a, Ezzeddine2020a, Francois2016a, frebel2010ApJ, Fulbright2000a, Gilmore2013a, Hansen_T2018a, Holmbeck2020a, Ishigaki2014a, jacobson2015, Ji2016a, Ji2016b, ji2019chemical, Lai2007a, Preston2000b, Roederer2010c, Sakari2018b, yong2013, Ou2020, Duong2019, Bensby2014, Afsar2018A, Afsar2012, Johnson2012, Boesgaard2011, Rasmussen2020, Navarrete2015, For2010, Reggiani2017, Peterson2013,  chiti2018chemical,ji2020detailed, roederer2016, marshall2019}).}
    \label{fig:elements}
    
\end{figure*}

We measured several $\alpha$-and iron-peak elements in our sample stars. This includes C, Mg, Si, Ca, Ti, Cr, and Ni. In Figure~\ref{fig:elements}, we compare our results with other known metal-poor stars in the halo (black points) and various UFDs (green squares). Triangles represent upper limit values.

The majority of our sample agrees well with abundance trends found among the halo and UFD population for each element. However, stars Gaia~DR3~4348475068024891776 {(purple cross)} and Gaia DR3 5729400267359655680 {(dark blue star)} are consistently found towards the edges of the distributions, with the latter having significantly lower values in Ti, Cr and Ni, among the lowest measured across all stars. In addition, our sample shows a significant spread in each element. This could indicate that the stars in our sample were born in different environments that underwent individual mixing events leading to some level of stochasticity.

{We note that Gaia DR3 4191531774709865984 (2MASS J20055457-0928302) displays low, subsolar [$\alpha$/Fe] ratios for Mg, Si and Ca. However, Ti has the usual enhancement of [Ti/Fe] $\sim$ 0.4. This general behavior has been seen in a few other "iron-enhanced metal-poor" stars (e.g., \citealt{yong2013, jacobson2015}) which signals early onset of type Ia supernova enrichment in the star's birth gas cloud, presumably in an early dwarf galaxy environment. Such origin scenario is supported by the supersolar Cr and Ni abundances.}

We validated our chemical abundances using two stars that had also been studied by \citet{Yong2021}. For SMSS J030634.26$-$750133.3 {(green square)}, we determined $\mathrm{T}_{\mathrm{eff}}$, $\log$ g, and [Fe/H] of 5148~K, $2.27\pm0.01$, and $-3.00\pm0.15$ respectively, compared to the \citet{Yong2021} values of 5075~K, 2.20, and $-3.14$. Our analysis of SMSS J030740.92$-$610018.8 {(cyan pentagon)} yielded measurements of 5014~K, $1.97\pm0.01$, and $-3.12\pm0.15$, in comparison to their results of 5025~K, 2.06, and $-3.09$.

When comparing abundance measurements with \citet{Yong2021}, we find that the majority of our measurements agree within $\pm0.15$\,dex for [X/Fe], with the largest difference being $-0.36$ for [Ba/Fe] in star SMSS J030740.92$-$610018.8. Overall, there is very close agreement between our measurements and those of \cite{Yong2021}.

\subsection{Neutron-capture element abundances and SASS star classification}

We begin by noting the discovery of a new ultra-metal-poor star, {Gaia DR3 5729400267359655680 (SMSS J090921.70$-$163821.1)}, with [Fe/H] $=-4.1$ and a hyper low Sr abundance of [Sr/H] =$-6.4$ which is the lowest [Sr/H] measurement known to date. Interestingly, the star also has a hyper low Ba abundance of [Ba/H] $=-6.0$, and unusually low iron-peak element abundances. This star was first identify at extremely metal-poor star in the SkyMapper survey, based on medium-resolution spectroscopy \citep{dacosta2019}. Further details on this star will be discussed in an independent paper.

\begin{figure}[ht]
    \centering
    \includegraphics[width=0.85\columnwidth]{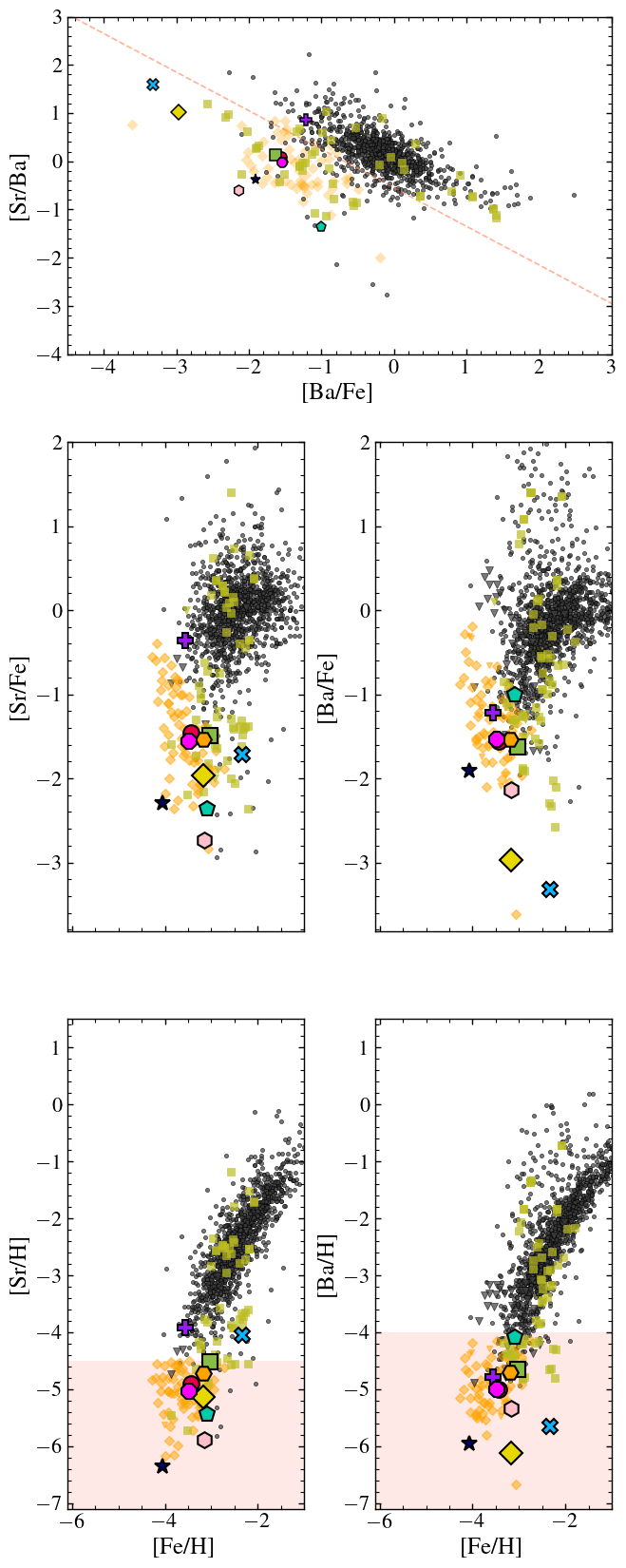}
    \caption{(\emph{Top}) Distribution of [Sr/Ba] relative to [Ba/Fe] in our sample, compared to halo (gray) and UFD (olive) stars. The orange dashed line  \citet{Andales} divides the tracks by halo stars and UFD stars. Nine of our ten stars fall within the UFD track (below the orange line). (\emph{Bottom}) Abundances of [Sr/H] and [Ba/H] as a function of [Fe/H] for our ten stars in comparison to UFD (olive squares) and halo stars (gray points). The pink shaded areas highlights the SASS criteria \citep{Andales}. Symbols are the same as in Figure~\ref{fig:elements}.}
    \label{fig:SrH_BaH}
\end{figure}

Regarding the entire sample, and building upon the analysis established in \cite{Andales}, we use the SASS definition to classify our ten stars. To summarize, SASS stars are defined by extremely low Sr and Ba values, namely [Sr/H]~$<-4.5$ and [Ba/H]~$<-4.0$, respectively. 
While the [Sr/H] vs [Fe/H] distribution indicates a separation (see Figure~\ref{fig:SrH_BaH}), the [Ba/H] distribution does not, leaving [Sr/H] as the main indicator. We here add an additional qualifier to the original definition: SASS stars should also adhere to [Sr/Fe] $< 0$ to ensure that the Sr and Ba values would truly reflect either early enrichment by one or few progenitors. 

SASS stars generally have [Sr/Ba] ratios that agree with those of UFD stars as can be seen in the top panel of Figure~\ref{fig:SrH_BaH} which shows the [Sr/Ba] ratios vs [Ba/Fe]. The orange line represents a visual boundary between the halo and UFD star populations. We caution that SASS-like [Sr/Ba] ratios, at face value, are found in many metal-poor halo stars \citep{roederer2017}; however these objects would not pass the [Sr/H]$<-4.5$ test for SASS membership.

Applying these SASS criteria to our sample of ten stars, eight are bona fide SASS stars. The remaining two stars, 2MASS~J20055457$-$0928302 {(blue diagonal cross)} and Gaia~DR3~4348475068024891776, only meet the [Ba/H] $< -4.0$ (and the [Sr/Fe] $<0$) requirement. Their Sr abundances are still low at [Sr/H] $=-4.11$ and $-3.92$, respectively.

In Figures~\ref{fig:SrH_BaH} (bottom panels), we present the distribution of neutron-capture element abundances for our sample of stars alongside known halo and UFD stars available from the literature. We choose to include our two non-SASS stars for completeness; however, they have been excluded from our remaining SASS population analysis. 

In addition, we also show here an additional 69 literature stars (including the stars from \citealt{Andales})  for a comprehensive analysis of the nature and origin of SASS stars. The SASS thresholds for Sr and Ba are indicated by the pink highlighted areas. We mark all SASS stars within this region with orange diamonds in the plot. 

For completeness, we note here there are another eight stars that have [Sr/H] $<-4.5$ (shows as small gray circles in Figure~3) but that are not classified as SASS stars  due to either high or absent Ba abundances.

\subsection{Kinematic and orbital properties of the sample}
\label{sec:kinematics}
{For completeness, and to learn more about the origins of our stars, we provide the kinematic information of our sample. The results of our kinematic analysis are presented in Table~\ref{tab:kinematics}.}

To construct their full space of motions, we use Gaia DR3 proper motions ($\mu_\alpha \cos{\delta},\mu_\delta$) and parallaxes ($\Bar{\omega}$) \citep{Gaia_DR3}, where $\Bar{\omega}$ is corrected following the procedure outlined in \cite{Lindegren_Parallax_2021}. 
{We use distances as determined in Section 3, and coordinates and radial velocities from Table~\ref{tab:kinematics}. } 

We furthermore adopted parameter values, described in \cite{Mardini2024}, as follows: The Sun is located at $R_{\odot} = 8.178\pm0.013$ kpc from \cite{Gravity_Collaboration2019}, $z_{\odot} = 20.8\pm0.3$ pc above the galactic plane, with a peculiar motion of $U_{\odot} = 11.1\pm0.72$ $\mathrm{kms}^{-1}$ \citep{Bennett2019}, $V_{\odot}= 12.24\pm0.47$ $\mathrm{kms}^{-1}$ and $W_{\odot} = 7.25\pm0.36$ $\mathrm{kms}^{-1}$ \citep{Schonrich2010}. Then $V_{\mathrm{LSR}}$ is taken as $220$ $\mathrm{kms}^{-1}$ \citep{Kerr1986}. 

{We use the
\texttt{The-ORIENT}\footnote{\url{https://github.com/Mohammad-Mardini/The-ORIENT}}  \citep{Mardini_2020,Mardini2022b} software to obtain stellar orbits, apocentric ($r_{\rm apo}$) and pericentric ($r_{\rm peri}$) radii, the maximum offset from the Galactic midplane ($Z_{\rm max}$), and eccentricity, defined as $e = (r_{\rm apo} - r_{\rm peri}) / (r_{\rm apo} + r_{\rm peri})$. The code utilizes a time-varying potential constructed from snapshots of the ILLUSTRIS TNG50 simulation \citep{2021MNRAS.507.4211E}, as described further in \cite{Mardini2024}.}

{Based on the Z$_{max}$, five stars appear to likely be associated with the outer halo and four with the inner halo. One, Gaia DR3 6726207068062588160, is likely a disk star.
It is also indicated that five stars are likely of an accretion origin as they have negative $L_{z}$ values. Moreover, the majority of the sample has high eccentricities ($e>0.5$) which indicates radial, plunging trajectories that bring stars from the outer halo into the inner Galaxy. Prograde stars with such extreme eccentricities can thus also be considered to have origins with disrupted satellites. This aligns with our earlier results from \citet{Andales}, whose $\sim60$ SASS stars exclusively showed retrograde motion, and hence, an accretion origin. }

As such, there is clear kinematic evidence for an accretion origin of SASS stars. In addition, their position in the [Sr/Ba] vs. [Ba/Fe] diagram overlap with those from UFD stars which, principally, also supports accretion. Therefore, our SASS criteria of [Sr/H] $<-4.5$ and [Ba/H] $<-4$ remain useful to select chemically primitive metal-poor stars with a likely accretion origin. {We note that we're agnostic to the type of progenitor system but candidates are UFDs and the earliest-formed classical dwarf galaxies \citep{yelland26}.}

\section{Strontium yields in the early universe} 

The low neutron-capture element abundances ([Sr/H] $<-4.5$ and [Ba/H] $<-4.0$) of the SASS stars imply that they must have formed in relatively primitive environments in the early universe with only one or few progenitor events and without significant neutron-capture element yields. This picture is supported by their low [Fe/H] $<-2.5$ and their likely accretion origins. Also, these stars do not appear to be associated with any known stellar streams, which implies their hosts to have been completely disrupted.

{Astrophysical sites responsible for neutron-capture element nucleosynthesis remains poorly understood and advances have rested entirely on theoretical predictions of yields. These are often limited to ratios rather than absolute values, and all depending on  progenitor properties and nuclear physics parameters for forward-modeling the nucleosynthesis processes. We here use the existence of these low Sr SASS stars provides an opportunity for trialing an empirical approach of inferring the progenitor supernovae yields directly from the observed abundances.}

\subsection{Strontium yields across multiple nucleosynthesis sites} 
\label{sr_yields}
From Figure~\ref{fig:SrH_BaH}, it can be seen that our eight new SASS stars with the lowest Sr abundances do not necessarily correlate with equivalently low Ba abundances. In fact, from the top panel, it can be seen that the [Sr/Ba] ratios for all SASS stars, including our sample, cover a large range from [Sr/Ba] $=-2.0$ to +1.6. As such, these stars likely formed under different nucleosynthetic conditions, implying distinct yields for Sr (and Ba) depending on the dominant enrichment process.

To constrain the multiple nucleosynthesis processes that likely resulted in the large [Sr/Ba] spread, in this Section, we investigate how early progenitor events may have contributed to the observed Sr abundances and whether there was any intrinsic scatter.

\par
We find and color-code four groups of [Sr/Ba] ratios within the SASS star population:
[Sr/Ba] $<-0.6$ (purple), 
$-0.6\leq$ [Sr/Ba] $<-0.2$ (blue), 
$-0.2\leq$ [Sr/Ba] $<0.5$ (yellow), and 
[Sr/Ba] $>0.5$ (red). 
The group with the lowest [Sr/Ba] ratio is the purple group with [Sr/Ba] $<-0.6$ (9 stars) which conservatively covers what could be expected to arise from the $s$-process (generally [Sr/Ba] $<-1.0$ at low-metallicity; \citealt{2008ARA&A..46..241S}). 
Given that most of the sample stars have metallicities around [Fe/H] $\sim-3$, we would not expect a significant contribution from any $s$-process operating in AGB stars to the enrichment of the respective birth gas clouds. However, a weak $s$-process operating in massive stars \citep{Chiappini2011, choplin2017} could be a possible alternative for producing such low [Sr/Ba] ratios. This indicates that their [Sr/Ba] ratios are even lower, with the lowest one being [Sr/Ba] =$-2.8$, suggesting that a weak $s$-process may have produced these more extreme ratios. Hence, these stars may well very similar to our stars purple group although we do not consider them further in this study.

We then have the blue group with $-0.6\leq$ [Sr/Ba] $<-0.2$ (31 stars), which is also the largest of the groups, with a [Sr/Ba] ratio commensurate with what is found in extremely metal-poor stars enriched by the main $r$-process ($\simeq-0.3$ to $-0.4$; \citealt{McWilliam1998}). Any r-process site operating in the early universe would need to be of a prompt nature in order to explain the low [Fe/H] of these stars. Fast merging NSM \citep{zevin2019} might be an option but so are exotic types of supernovae (e.g., \citealt{siegel19}). 
The yellow group (27 stars) spans $-0.2\leq$ [Sr/Ba] $<0.5$. At present, no known nucleosynthesis process produces [Sr/Ba] $\sim$ 0 but we note that this corresponds to about five times more Sr than Ba production which could be interpreted as a mild limited $r$-process. Alternatively, it could be a combination of an r-process along with enrichment from one or multiple limited r-processes. 
Finally, the red group with [Sr/Ba] $>0.5$ (9 stars) corresponds to the limited $r$-process which is thought to produce progressively lower heavier neutron-capture element abundances, such as Ba, compared to lighter ones. The benchmark limited-$r$ star HD122563 \citep{honda06} has [Sr/Ba] = +0.76. Electron-capture supernovae or core-collapse supernova are likely the main site of the limited $r$-process \citep{wanajo2011, arcones2013}. 

The large spread in [Sr/Ba] ratios clearly suggests that no single Sr yield per event can adequately describe the full SASS star population. 
We thus begin to explore early neutron-capture element enrichment by assuming that there must be fixed (heavy element) yields per event for each [Sr/Ba] group, and that SASS stars will have formed from gas enriched by $\ge1$ nucleosynthesis event. 


We then choose to adopt [Sr/H] values, one for each of the [Sr/Ba] groups, that each could reflect a single Sr yield per event and approximately match the average of the lowest several observed [Sr/H] abundances in each group. We elaborate further below on our procedure of determining these minimum Sr values for each group.

We convert the [Sr/H] abundances into Sr yields in solar masses using the following equation
\begin{equation}
    \mathrm{Y}\,[M{_\odot}] = A_{\mathrm{Sr}} M_{\mathrm{gas}} 10^{\left(\rm{[Sr/H]}_\star +~\log \epsilon(\mathrm{Sr})_{\odot} - 12\right)},
    \label{eq:yield}
\end{equation}

where we assume a fixed gas mixing mass of M$_{\mathrm{gas}}$ = $10^5$\,M$_\odot$, and $A_{\mathrm{Sr}}$ is the atomic mass of Sr (88), and 
$\log \epsilon(\mathrm{Sr})_{\odot} - 12 = -9.13$ represents the solar Sr abundance with the respect to hydrogen on a scale of 12. 

This results in the following:
$[\mathrm{Sr/H}]=-6$ corresponds to $Y_{\rm Sr}=6.52\times10^{-9}\,{\rm M}_{\odot}$ (purple group);
$[\mathrm{Sr/H}]=-5.75$ to $1.16\times10^{-8}\,{\rm M}_{\odot}$ (blue group);
$[\mathrm{Sr/H}]=-5.42$ to $2.48\times10^{-8}\,{\rm M}_{\odot}$ (yellow group);
and $[\mathrm{Sr/H}]=-4.93$ to $7.66\times10^{-8}\,{\rm M}_{\odot}$ (red group).

We order these groups by their [Sr/Ba] ratio but note that the red group has lower Sr yield than than the yellow group. This is likely due to the red group only having nine stars. Regardless, the yield prediction for the red group is the most uncertain value.

Sr is preferred over Ba as our reference element because it is produced in significant amounts across various neutron-capture element production channels, including the main r-process, limited r-process, and massive-star weak s-process. 
Ba production, by contrast, depends much more on the specific process, especially for the limited r-process which suppresses it relative to that of lighter neutron-capture elements such as Sr \citep{honda06}. 
However, our yields can be considered largely agnostic to the nucleosynthesis process that actually produced the Sr -- the only assumption we make here is that it occurred in a prompt enrichment source, such as a type of supernova, given the very low metallicities of our stars. 

We also assume, per group, that both Sr and Ba observed in a given star were produced in the same progenitor event, and that its birth gas clouds was enriched by identical progenitor events (i.e., with approximately the same [Sr/Ba] value) rather than different yields or a combination of different enrichment sites.

Using these single Sr yields per group at face value, for completeness, we find the corresponding single Ba yield for each group, based on average/approximate [Sr/Ba] group values: 
[Sr/Ba] = $-1.1$ with [Sr/H] = $-6$ corresponds to [$\mathrm{Ba/H}]=-4.9$ and $Y_{\rm Ba}=2.61\times10^{-8}\,{\rm M}_{\odot}$ (purple group);
$[\mathrm{Sr/Ba}]=-0.4$ with [Sr/H] = $-5.75$ to $[\mathrm{Ba/H}]=-5.35$ and 
$Y_{\rm Ba}=9.26\times10^{-9}\,{\rm M}_{\odot}$ (blue group);
$[\mathrm{Sr/Ba}]=0.15$ with [Sr/H] = $-5.42$ to $[\mathrm{Ba/H}]=-5.57$ and $Y_{\rm Ba}=5.58\times10^{-9}\,{\rm M}_{\odot}$ (yellow group);
and $[\mathrm{Sr/Ba}] = 1.0$ with [Sr/H] = $-4.93$ to $[\mathrm{Ba/H}]=-5.93$ and $Y_{\rm Ba}=2.44\times10^{-9}\,{\rm M}_{\odot}$ (red group).

{We note that these specific Sr yields derived for each group depend on the adopted Sr/Ba separation. However, choosing just two groups would impact the respective yields at a level of 0.2-0.3\,dex only. As such, we assume that our group choices do not significantly affect any conclusions.}

\begin{figure}
    \centering
    \includegraphics[width=\columnwidth]{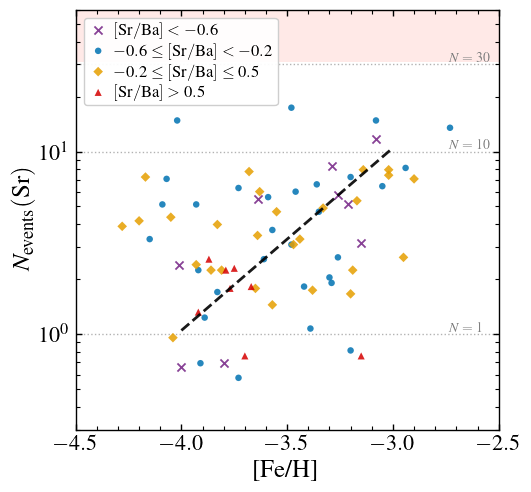}
    \caption{ Number of supernova events ${\rm N}_{\rm events(Sr)}$, based on individual group yields between [Sr/H] = $-6$ and $-4.99$, as a function of metallicity [Fe/H]. SASS stars are color-coded according to their observed [Sr/Ba] ratio. The black dashed line represents the Sr-Fe co-production relation, shifted to match that of the lowest [Sr/Ba] group (purple).}
    \label{fig:combined_SN}
\end{figure}

Based on the four Sr yields, we then calculate the number of events that contributed to the observed abundance of Sr of each star, by dividing the observed [Sr/H] abundances by $Y_{\rm Sr}$ to obtain the number of $\mathrm{N}_{\rm events(Sr)}$ that enriched the birth gas cloud. 
By design, the majority of the SASS stars have few progenitor events, around 1-10,  as shown in Figure~\ref{fig:combined_SN}. This is commensurate with the fact that our sample is within the extremely metal-poor regime. Since the SASS criterion is [Sr/H]$\leq-4.5$, we also note for completeness that this imposes a principal cut off of $\sim30$ progenitor events. 

The stars in the yellow and red groups tend to have less than 10 progenitor events. In contrast, stars in the two lowest [Sr/Ba] groups (purple and blue) showcase the full range of possible progenitor events given our [Sr/H] cuts. This aligns with the common assumption that extremely metal-poor stars with [Fe/H] $\lesssim-3$ all form from gas enriched by few, perhaps less than 10, progenitor events. We note that HE~1327$-$2326 (with [Fe/H] $-5.4$) is not included in the SASS sample given that this star has an unusually \textit{high} abundances of [Sr/Fe] = 1.2 \citep{frebel2005nucleosynthetic} which fails our SASS criteria. 
This reduces our final total sample to 76 SASS stars which includes our eight new SASS stars.

In order to actually determine the minimum [Sr/H] values for each group, we had first consider the Fe abundances of SASS stars. Given that the Fe observed in each star is expected to also have been produced in the same progenitor event than the Sr and Ba but independently of the heavy elements, we here assume, for simplicity, a linear relation (``co-production'') between the number of events (a proxy for the Sr abundances) and the observed Fe abundances, for each of the four [Sr/Ba] groups.

{However, the stars do not display any obvious correlation between number of supernovae and measured Fe abundances, yet we still associate one co-production relation. 
By fitting a linear function to the median of each group, we find moderate differences in [Sr/H], with stars in the yellow group having, on average, higher [Sr/H] abundances, compared to the e.g., purple group. The purple group's co-production relation has the lowest median, corresponding to the [Sr/H] $-6$ yield.} 

{For each of the groups, we take the lowest observed [Sr/H] value as the respective single yield. They vary by a few tens of a dex, with the [Sr/H] $-6$ yield being the lowest one (purple group). Calculating the number of supernovae for all [Sr/Ba] groups then matches the median of the purple group. This is shown in Figure~\ref{fig:combined_SN}. }

{As can be gleaned from the Figure, there is significant scatter in each group. Not only have we sorted all our sample stars into four different [Sr/Ba] groups, each group on its own shows significant scatter. This implies that no one individual site (associated with a specific [Sr/Ba] ratio) could adequately describe the full population of SASS stars.} 

{While we have here assumed that enrichment would occur by identical supernovae in each [Sr/Ba] group, the scatter clearly suggests that this kind of enrichment is more complex. For each group, it could in fact be a mixture of progenitor sites, or the Fe yields for each supernovae could be variable to within a factor of 10, or both.}

\subsection{Estimating the range and scatter of Sr yields per nucleosynthesis site}
\label{sect:gas_mass}

In the previous Section, for the stars in the four [Sr/Ba] groups, we have adopted distinct Sr yields to establish how many progenitor events contributed to their observed Sr abundances. 
{Since these stars formed from gas experiencing very few progenitor events, we assume one nucleosynthesis process to dominate within each group. The residual scatter within each group then simply reflects a range of the associated yields.}

Scaling the observed abundances with their respective yield aligned the groups (see Figure~\ref{fig:combined_SN}) and somewhat reduced the overall scatter of the SASS sample by a factor of 2-3. Yet, as can be seen in  Figure~\ref{fig:combined_SN}, the stars still scatter significantly around the Sr-Fe co-production relation. {We interpret this scatter as variations in the respective Sr yield.}

{We note that the observed scatter might very well arise in two other ways: 1) A given nucleosynthesis process there could easily be variations in the Sr yield, 2) The gas clouds could also have been enriched by more than one nucleosynthesis process. The former possibility is somewhat accounted for in our choice of [Sr/Ba] bin size for our four groups. However, as mentioned above, for simplicity, we do not account for or consider the latter possibility of mixed enrichment.}

\begin{figure*}
    \centering
    \includegraphics[width=\linewidth]{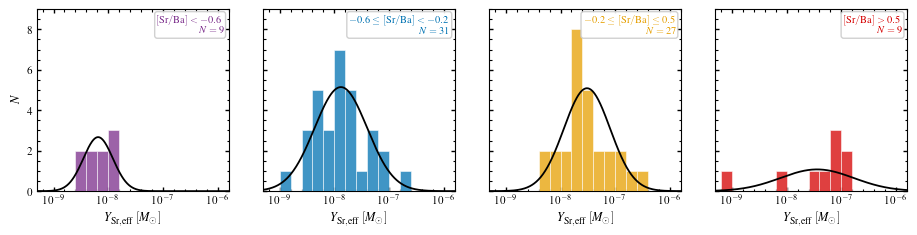}
    \caption{Distribution of the scatter in the Sr yields relative to the Sr-Fe co-production relation (see Figure~4)  for the purple, blue, yellow and red groups.}
    \label{fig:yield_dist}
\end{figure*}

In Figure~\ref{fig:yield_dist}, we show the distributions of Sr yields for each [Sr/Ba] group. 
 
They reflect the range of yields possible for each distinct underlying nucleosynthesis process of each group.
The blue group displays the largest and most well-established spread from $1.4\times10^{-9}$ to $\sim2\times10^{-7}$\,$M_\odot$, pointing to a significant continuum of Sr yields associated with (main) r-process nucleosynthesis. In contrast, the purple and red groups exhibit the smallest spread (approximately an order of magnitude) but they also has significantly fewer stars.
We note that the peaks of these distributions do not exactly match the adopted single yield per group given low number statistics and the fact that we used the medians of the four groups to determine the single yields (see Section~4). 

Overall, the spreads of these distributions indicate that the Sr yields for each group have a typical spread of a factor of no more than 100 in total, and with a full width at half maximum (FWHM) of less than or about a factor of 10. 
While a single fixed yield per [Sr/Ba] group (the peak value) will never be sufficient to describe the apparently significant range of variations per nucleosynthesis process that led to the formation of these metal-poor stars, the surprisingly narrow FWHM suggest that any adopted Sr yield close to our single Sr yield (per group) would be good to within a factor of a few.

However, an outstanding question remains as to how reasonable, on an absolute level, our initial choices for the single Sr yields for each [Sr/Ba] group were, given that we could have equally chosen even lower values as baseline yields. 

We argue that the lower end of any of these Sr yield distributions should be no more than (approximately) the lowest observed [Sr/H] abundances in each group to ensure at least one progenitor nominally produced what is observed (i.e., it cannot be less than one event). Moreover, a distribution that would have its lower end at an even (significantly) lower yield would imply that a corresponding number of additional progenitor events would be required to interpret the observed Sr abundances; however, that would also impact the general [Fe/H] abundance interpretation, and likely (misleadingly) suggest that these extremely and ultra-metal-poor star would have had many more progenitors than what their [Fe/H] would canonically suggest.

{We note that throughout this study, we employed a fixed birth gas cloud mass of $10^5\,M_\odot$ for calculating the dilution of all observed abundances, since Sr yield and cloud mass choice are degenerate. The only available variable is adopting a different gas mass altogether, which then just shifts the yields accordingly. In this context, we note that, for simplicity, we assume the mass of the birth gas cloud to be the same for all stars in all groups. }

{We thus conclude, despite the gas cloud mass assumption, that our adopted Sr yields for each [Sr/Ba] group are reasonable estimates of the \textit{representative} Sr yields of their respective underlying nucleosynthesis process (i.e., for each [Sr/Ba] group), within a factor of a few. Consequently, this approach also leaves room for a factor of a few variations in the assumed gas mass even though we do not explicitly address this further.} 

{Along the same lines, we note that our adopted single Sr yields could principally be adjusted (independently for each group) either because of 1) a different gas cloud mass choice or 2) because of our choice for the single yield (i.e., which lowest observed [Sr/H] is adopted). Neither choice would affect the scatter as seen in Figure~4 or the corresponding yield distributions (see Figure~5). Any changes would simply shift the distributions but not alter their widths.}

{The total width reflecting yield variations of a factor of 100 or less may then be compared with results from other, similar exploratory studies, or be adopted by future theoretical nuclear astrophysics or chemical evolution models.}


Overall, this exploration shows that, for now, SASS stars reflect a range of Sr yields from multiple types of nucleosynthesis events that led to their observed heavy element abundances. Additional discoveries of SASS stars, in particular, stars in the red group with $[\mathrm{Sr/Ba}] > 0.5$ could provide further insights into the yields and possible variations of the underlying heavy element nucleosynthesis process (i.e., for each [Sr/Ba] groups). This would enable further refinements of the yields of Sr, and potentially offer the opportunity to establish yields of other heavy elements.

\section{Discussion and Summary}

In this paper, we present a chemical abundance analysis for a new sample extremely metal-poor halo stars, of which eight meet the SASS criteria defined as [Sr/H] $< -4.5$ and [Ba/H] $< -4$. Given their low iron and extremely low neutron-capture-element abundances, these SASS stars are thought to have been enriched by very few, or even just a single, early supernova event. Therefore, SASS stars are important tracers for the contributions of the various heavy-element nucleosynthesis processes operating in the early universe, what their yields were and what their intrinsic scatter was. Our study has investigated these topics, and the key results are as follows:

\begin{enumerate}
  \renewcommand{\labelenumi}{(\roman{enumi})}
  \item Of our eight new SASS stars, we find 2MASS J090921.7$-$163821.1 to be a hyper neutron-capture element poor star with [Sr/H] $= -6.4\pm0.1$, which is also the lowest [Sr/H] abundance measurement within the current literature. 
  
  \item The full SASS star sample covers a very large range of [Sr/Ba] ratios, from $-1.6<$ [Sr/Ba] $<+2.0$, which suggests that multiple distinct nucleosynthetic processes must have operated in the early universe.

  \item We assume a co-production of Sr and Fe, and place such relation at the bottom ends of the [Sr/H] abundance distributions for each [Sr/Ba] group. 
  The relative offsets between each group and the lowest (purple) group are used to estimate group-specific Sr yields of the corresponding progenitor nucleosynthesis events. The resulting single-event Sr yields are [Sr/H] $=-6.0$ (purple group), [Sr/H] =$-5.75$ (blue group), [Sr/H] =$-5.42$ (yellow group), and [Sr/H] =$-4.93$ (red group), respectively. Using average [Sr/Ba] from each group, we also obtain the matching Ba yields of [Ba/H] =$-4.9$ (purple group), [Ba/H] =$-5.35$ (blue group), [Ba/H] =$-5.57$ (yellow group), and [Ba/H] =$-5.93$ (red group). In Section~\ref{sr_yields}, we also provide these yields in solar masses.

  \item Invoking these four Sr nucleosynthesis yields, the SASS stars formed from gas enriched by 1–10 progenitor events. This aligns with expectations of the extremely and ultra-metal-poor nature. 

  \item A kinematic analysis of our sample supports the notion that all the SASS stars likely have an accretion origin. As such, the number of progenitor nucleosynthesis events characterizes the early star forming environments, such as in early analogs of the surviving UFDs or classical dwarf galaxies \citep{yelland26}.     
  
  \item Despite using a different single yield for each group, scatter around the expected Sr-Fe co-production relation remains. This suggests each nucleosynthesis site produces a range of Sr yields rather than a fixed value. The inferred yield distributions span a factor of $\leq100$ in total, but their FWHM of less than a factor of $\sim10$ suggest our single yield per group remains a reasonable representative value to within a factor of a few.

\end{enumerate}

Our findings strongly indicate that the earliest neutron-capture element production was driven by a variety of nucleosynthetic processes with distinct, but perhaps overlapping, yields, as evidenced by the wide [Sr/Ba] spread across the SASS population. The spread of the inferred Sr yield distributions further demonstrates that early enrichment environments were far from uniform.

{Looking ahead, our new approach here allows us to apply the individual Sr yields to other environments such as} surviving UFDs which are thought to have been enriched by one or very few neutron-capture element producing events. After all, the original host systems of SASS stars would have been early analogs of the surviving UFDs.
For the UFD Grus I, there is one stars with [Sr/Ba] $=-0.42$ and [Sr/H] = $-4.57$ \citep{ji2019chemical} that aligns with our SASS criterion. This suggests early enrichment related to the r-process and produced by a dozen progenitor events. However, a second star with [Sr/Ba] $=-1.12$, [Sr/H] = $-4.56$ and a higher Ba abundances of [Ba/H] $=-3.44$ is also present in Grus I. This suggests the early nucleosynthesis was varied, possibly also driven by a weak s-process that would have operated massive supernovae. A holistic view of what SASS stars and other stars with generally low [Sr/H] and [Ba/H] is required when trying to understand the full scope of early heavy element enrichment in a given system or region.

{We also attempt to applying this framework to the r-process enriched UFD Reticulum II, because it also contains two ``normal'' stars with [Sr/H] $<-4.24$ and $<-3.54$  besides the r-process-enhanced stars \citep{ji2016complete}. } 
Unfortunately, since only upper limits are measured in those two stars, we cannot distinguish the type of early progenitor events. But irrespective of which of Sr yields one might pick, the observed abundance of the star with [Sr/H] $<-4.24$ suggests that no more than a couple dozen events enriched the early Reticulum II. Presumably, other heavy element processes operated at early times (limited r-process, weak s-process; depending on what the [Sr/Ba] might be) before a prolific r-process enrichment took place that led to the formation of the r-process stars present in Reticulum II.

{Taken altogether, our approach of investigating the sample of 76 SASS stars, along with defining different Sr yields of the progenitor events -- presumed to be early supernovae of some kind -- provides a new approach of investigating heavy element nucleosynthesis in the era of early star and galaxy formation. The SASS sample shows that multiple distinct neutron-capture-element-producing sites are likely required, and that these collectively give rise to the large spread in observed [Sr/Ba] ratios. }

Our approach also provides a starting point for additional theoretical nucleosynthesis calculations to reproduce our ``observed-inspired'' Sr yields for each of the four nucleosynthesis sites, as well as for improving chemical evolution modeling of the first/early galaxies and the onset of the hierarchical assembly.

\begin{acknowledgements} 

A.F., X.O., and A.Y. acknowledge support from NSF-AAG grant AST-2307436. 
F.X. and J.B. acknowledge  support  from  the  MIT  UROP  program. M.M. and A.F. acknowledge support from the MISTI-Jordan Global Seed Fund, A.F. acknowledges support from the PCLB Foundation. 

This work has made use of data from the European Space Agency (ESA) mission
{\it Gaia} (\url{https://www.cosmos.esa.int/gaia}), processed by the {\it Gaia}
Data Processing and Analysis Consortium (DPAC,
\url{https://www.cosmos.esa.int/web/gaia/dpac/consortium}). Funding for the DPAC
has been provided by national institutions, in particular the institutions
participating in the {\it Gaia} Multilateral Agreement.
This research has made use of the SIMBAD database,
operated at CDS, Strasbourg, France.

\end{acknowledgements}

\bibliographystyle{aasjournal}

\end{document}